\documentstyle[12pt]{article}

\catcode`@=12
\begin{document}
\thispagestyle{empty}
\begin{flushright} UCRHEP-T187\\May 1997\
\end{flushright}
\vspace{0.5in}
\begin{center}
{\Large\bf Seeding of Strange Matter with New Physics\footnote{Talk given 
at the {\it International Symposium on ``Strangeness in Quark Matter 1997"} 
(Santorini, Greece) April, 1997, to be published in J. Phys. G} \\}
\vspace{1.8in}
{\bf Ernest Ma\\}
\vspace{0.3in}
{\sl Department of Physics\\}
{\sl University of California\\}
{\sl Riverside, California 92521, USA\\}
\vspace{1.5in}
\end{center}
\begin{abstract}\
At greater than nuclear densities, matter may convert into a mixture of 
nucleons, hyperons, \underline {dibaryons}, and \underline {strangelets}, 
thus facilitating the formation of strange matter even before the onset of 
the quark-matter phase transition.  From a nonstrange dibaryon condensate, 
it may even be possible to leapfrog into strange matter with a certain new 
interaction, represented by an effective six-quark operator which is 
phenomenologically unconstrained.
\end{abstract}
\newpage
\baselineskip 24pt
 
\section{Introduction}

The various forms of matter which may exist depend on external conditions 
such as temperature and pressure.  Our knowledge of particles and nuclei 
comes mainly from experiments near zero temperature and zero pressure, 
although current and future heavy-ion colliders will probe further into 
these regimes.  A very important question to be answered is whether 
strange quark matter\cite{1,2} is stable or metastable under ordinary 
conditions.  We know that at zero pressure, two-flavor ($u$ and $d$) 
quark matter must be highly unstable, or else we would not have stable 
nuclei.  We also know that particles containing strangeness (hyperons) 
are unstable and decay weakly.  However, it is still possible for 
strange quark matter (containing $u$, $d$, and $s$ quarks) to be stable 
or metastable, and be undetected so far because it is not easy to make. 
After all, iron is the most stable of nuclei, but it only exists because 
it is made under extreme conditions in the interior of ordinary stars. 
A possible scenario for strange quark matter is that it is made inside 
massive neutron stars where the core densities are high enough to pass 
through the quark-matter phase transition.  It has also been 
suggested\cite{3} that some neutron stars in low-mass x-ray binaries can 
accrete sufficient mass to undergo a phase transition to become strange 
stars, for a possible explanation of the observed $\gamma$-ray bursters.

In the following, I propose the possibility that some forms of strange 
matter are already present inside neutron stars at densities far below 
that of the quark-matter phase transition.  This seeding of strange 
matter may be possible just from a proper treatment of the physical 
conditions for stability and equilibrium at high pressure, or more 
speculatively from a new interaction, represented by an 
effective six-quark operator which is phenomenologically unconstrained.

\section{Strangeness Content of Neutron Stars}

It has been known\cite{4} for a long time that at two to three times 
nuclear density, hyperons such as $\Lambda$ and $\Sigma$ particles 
may appear inside neutron stars.  At last year's Strangeness meeting in 
Budapest, two detailed analyses were presented\cite{5,6}.  To obtain 
these results, one must first know that hyperons exist and then have a 
model describing their interactions with nucleons, electrons, and 
neutrinos, {\it etc.}  Suppose one does not know about hyperons, but 
instead there is evidence for (nonstrange) dibaryons, then a model with 
dibaryons should be considered inside neutron stars.  Just such a 
possibility was recently considered\cite{7} and the conclusion was that 
a Bose condensate of dibaryons can occur in nuclear matter before the 
quark-hadron phase transition.

The lesson we learned from the above is that if we do not know the proper 
degrees of freedom relevant to the physical conditions of the external 
environment, then we cannot hope to arrive at the correct conclusion in 
that particular case.  There are things that we do know, such as the 
existence of hyperons and how they interact under laboratory conditions, 
but there are also many things that we do not know, such as whether 
dibaryons are stable or metastable or not, and perhaps more importantly, 
whether they can be stable in a high-pressure environment.  Our ignorance 
by itself cannot justify their omission in a full analysis of what may 
happen in a neutron star.  It would be very interesting to include 
\underline {both} hyperons and dibaryons, and perhaps even some 
\underline {strangelets}, and postulate interactions among them such as
\begin{eqnarray}
\Lambda \Lambda &\rightarrow& H ~(A=2, S=-2), \\ 
H H &\rightarrow& T ~(A=4, S=-4), \\ 
H T &\rightarrow& S^* ~(A=6, S=-6),
\end{eqnarray}
where $H$ is the $(uds)^2$ dibaryon\cite{8}, and $S^*$ is the quark-alpha 
particle\cite{9}.  It is important to realize that even if the $H$, $T$, 
and $S^*$ particles do not exist at zero pressure, their dynamics may be such 
that the above processes have to be included.  These quark-alphas should  
then be locally stable\cite{10} and may coalesce into bulk strange matter 
within this environment.

\section{New $\Delta$S = 3 Interaction}

If only nonstrange dibaryons form, then a nudge from some new physics may be 
advantageous for the formation of strange matter.  This kind of speculation 
may or may not be valid depending on the answers to the following two 
questions: (1) What kind of new physics is required? (2) Is it allowed by 
current experimental data?

Weak interactions change a $d$ quark into an $s$ quark via the $W$ boson 
according to the effective four-quark interaction
\begin{equation}
{\cal L}_{eff} = {G_F \over \sqrt 2} \sin \theta_C \cos \theta_C ~
\bar s \gamma_\mu (1 - \gamma_5) u ~ \bar u \gamma^\mu (1 - \gamma_5) d + h.c.
\end{equation}
which has the selection rule $\Delta$S = 1.  This is how two-flavor quark 
matter may convert into strange matter at sufficiently high densities.

Consider now the effective four-quark operator $(\bar s d)^2$ which 
has the selection rule $\Delta$S = 2.  Its magnitude is very much 
suppressed phenomenologically by the smallness of the $K_L - K_S$ mass 
difference and is approximately given by $G_F^2 m_c^2 
\sin^2 \theta_C \cos^2 \theta_C/16 \pi^2$.

What about a $\Delta$S = 3 interaction?  If it comes from the known 
weak interaction, it will of course be totally negligible.  However, 
new interactions may generate effective operators such as
\begin{equation}
{\cal L}'_{eff} = {1 \over M^5} (\bar s_R \bar s_R \bar s_R)_8 
(d_R d_R d_R)_8 + h.c.,
\end{equation}
where the subscript 8 refers to the color octet combination of the three 
quark triplets inside the parentheses.  This operator may not be 
negligible as shown below.

The effective interaction (5) may cause transitions such as
\begin{equation}
\Xi^0 (uss) \rightarrow n (udd) + K^0 (d \bar s),
\end{equation}
but since $1315 < 940 + 498$, it is kinematically forbidden.  In fact, the 
only allowed decay from (5) is
\begin{equation}
\Omega^- (sss) \rightarrow n (udd) + \pi^- (d \bar u),
\end{equation}
where $1672 > 940 + 140$.  The experimental upper limit on this branching 
fraction is about $10^{-4}$.  [There are only a total of 
$1.2 \times 10^4$ observed $\Omega^-$ events in the world.]  Since the 
$s$ quarks in $\Omega^-$ form a color singlet, whereas the operator (5) 
involves a color octet, two gluons must also be emitted for the proper 
rearrangement of the color wavefunctions.  Hence the amplitude for 
$\Omega^- \rightarrow n \pi^-$ is proportional to
\begin{equation}
{\alpha_S \over M^5} |\psi(0)|^2 \sim {\alpha_S \over {M^5 \pi R^3}},
\end{equation}
where $R \sim 1$ fm.  Comparing this against the dominant decays of 
$\Omega^-$ to $\Lambda K^-$ and $\Xi^0 \pi^-$, the bound
\begin{equation}
\left( {\alpha_S \over {M^5 \pi R^3 G_F \sin \theta_C \theta_C}} \right)^2 
< 10^{-4}
\end{equation}
can then be obtained.  For $\alpha_S \sim 0.3$, this implies $M > 7.9$ GeV.

Actually, because the $\Delta$S = 2 interaction is so small, a bound can 
be obtained also by tying one pair of $d$ and $s$ quarks in (5) to simulate 
an effective $(\bar s d)^2$ operator.  Hence
\begin{equation}
{1 \over M^5} {{m_s m_d m_c^2} \over {(16 \pi^2)^2 M_W^2}} \sin \theta_C 
\cos \theta_C m_K < {{G_F^2 m_c^2 \sin^2 \theta_C \cos^2 \theta_C} \over 
{16 \pi^2}}
\end{equation}
is required, which implies $M > 1.9$ GeV for $m_s = 150$ MeV and 
$m_d = 10$ MeV.

The above phenomenological analysis shows that the new physics mass scale 
is not very much constrained.  On the other hand, any specific model for 
realizing (5) will require the existence of new particles whose interactions 
with the $d$ and $s$ quarks would be such that they should be produced in 
current high-energy hadron colliders if their masses are of order 100 GeV 
or less.  For example, let there be an exact discrete $Z_3$ symmetry so 
that under $SU(3)_C \times U(1)_Q \times Z_3$,
\begin{equation}
d_R \sim (3, -{1 \over 3}, \omega), ~~~ s_R \sim (3, -{1 \over 3}, \omega^2), 
\end{equation}
where $\omega^3 = 1$.  Add two exotic scalar particles
\begin{equation}
Q_d \sim (6, -{2 \over 3}, \omega^2), ~~~ Q_s \sim (6, -{2 \over 3}, \omega),
\end{equation}
and one exotic fermion
\begin{equation}
\psi_8 \sim (8, -1, 1).
\end{equation}
Then the desired interaction (5) is obtained.  It should be noted that the 
above scenario conserves baryon number whereas a similar one proposed 
earlier does not\cite{11,12}.  See Fig.~2 of Ref.~[12].  The proposed 
exotic particles have strong interactions and should be copiously 
produced in high-energy hadron colliders, but their detection may be 
difficult since they may decay only into $d$ and $s$ quarks.

\section{Conclusion and Outlook}

At greater than nuclear densities, matter may convert into a mixture of 
nucleons, hyperons, \underline {dibaryons}, and \underline {strangelets}, 
thus facilitating the formation of strange matter even before the onset of 
the quark-matter phase transition.  The fact that we have very little 
laboratory information on $\Lambda \Lambda$ interactions does not preclude 
an important role for such and other processes inside neutron stars. 
This should be studied further in the future.  On the speculative side, 
it may even be possible to start from a nonstrange dibaryon condensate 
and leapfrog into strange matter with the six-quark 
effective interaction $(\bar s \bar s \bar s)_8 (d d d)_8$, whose magnitude 
is not very much phenomenologically constrained.
\newpage
\begin{center} {ACKNOWLEDGEMENT}
\end{center}

I thank the organizers of 
{\it SQM97} for their great hospitality.  This work was supported 
in part by the U. S. Department of Energy under Grant No. DE-FG03-94ER40837.

\bibliographystyle{unsrt} 

\end{document}